# Growth and form of the mound in Gale Crater, Mars: Slope-wind enhanced erosion and transport

**Edwin S. Kite[1], Kevin W. Lewis,[2] Michael P. Lamb[1]**

[1]Geological & Planetary Science, California Institute of Technology, MC 150-21, Pasadena CA 91125, USA. [2]Department of Geosciences, Princeton University, Guyot Hall, Princeton NJ 08544, USA.

**Abstract**: Ancient sediments provide archives of climate and habitability on Mars (*1,2*). Gale Crater, the landing site for the Mars Science Laboratory (MSL), hosts a 5 km high sedimentary mound (*3-5*). Hypotheses for mound formation include evaporitic, lacustrine, fluviodeltaic, and aeolian processes (*1-8*), but the origin and original extent of Gale's mound is unknown. Here we show new measurements of sedimentary strata within the mound that indicate ~3° outward dips oriented radially away from the mound center, inconsistent with the first three hypotheses. Moreover, although mounds are widely considered to be erosional remnants of a once crater-filling unit (*2,8-9*), we find that the Gale mound's current form is close to its maximal extent. Instead we propose that the mound's structure, stratigraphy, and current shape can be explained by growth in place near the center of the crater mediated by wind-topography feedbacks. Our model shows how sediment can initially accrete near the crater center far from crater-wall katabatic winds, until the increasing relief of the resulting mound generates mound-flank slope-winds strong enough to erode the mound. Our results indicate mound formation by airfall-dominated deposition with a limited role for lacustrine and fluvial activity, and potentially limited organic carbon preservation. Morphodynamic feedbacks between wind and topography are widely applicable to a range of sedimentary mounds and ice mounds across the Martian surface (*9-15*), and possibly other planets.

**Main text:** Most of Mars' known sedimentary rocks are in the form of intra-crater or canyon mounded deposits like the Gale mound (*2,9*), but identifying the physical mechanism(s) that explain mound growth and form has proved challenging, in part because these deposits have no clear analog on Earth. The current prevailing view on the formation of intra-crater mounds is that sedimentary layers (i.e., beds) completely filled each crater at least to the summit of the present-day mound (*2*). Subsequent aeolian erosion, decoupled from the deposition of the layers, is invoked to explain the present-day topography (*8,9*). If the sedimentary rocks formed as subhorizontal layers in an evaporitic playa-like setting, then $>>10^6$ $km^3$ must have been removed to produce the modern moats and mounds (*8,9*). These scenarios predict near-horizontal or slightly radially-inward dipping layers controlled by surface or ground water levels. To test this, we made 81 bed-orientation measurements from six one-meter-scale stereo elevation models, finding that layers have shallow but significant dips away from the mound center (Figures 1,S1), implying 3-4 km of pre-erosional stratigraphic relief. In addition, measurements of a dark-toned, erosionally resistant marker bed that varies in elevation by >1 km indicates beds are not planar. Postdepositional tilting with this pattern is unlikely (Supplementary Text), and these measurements permit only a minor role (*6*) for deposition mechanisms that preferentially fill topographic lows (e.g., playa, fluviodeltaic or lacustrine sedimentation), but are entirely consistent with aeolian processes (Figure S2). This suggests the mound grew with its modern elliptical shape, and that the processes sculpting the modern mound also molded the growing mound.

Mars is a windy place: saltating sand-sized particles are in active motion on Mars, at rates that predict aeolian erosion of bedrock at 10-50 μm/yr (*16*). Aeolian erosion of rock has occurred within the last ~1-10 Ka (*17*) and is probably ongoing. The inability of General Circulation Models (GCMs) to reproduce these observations shows that small-scale winds, not the regional-to-global winds resolved by GCMs, are responsible for saltation. Because of Mars' thin atmosphere, slope winds dominate the circulation in craters and canyons (*18*). Downslope-oriented yardangs, crater statistics, exposed layers, and lag deposits show that sedimentary mounds in Gale and Valles Marineris are being actively eroded by slope winds (*9*). Slope-enhanced winds also define both the large-scale and small-scale topography and stratigraphy of the polar layered deposits (*13,14*), and radar sounding of intracrater ice mounds near the north polar ice sheet proves that these grew from a central core, suggesting a role for slope winds (*15*). Most of the ancient stratigraphy explored by the Opportunity rover is aeolian (*19*), and aeolian deposits likely represent a volumetrically significant component of the sedimentary rock record, including within the strata of the Gale mound (*4*). Evidence for fluvial reworking within sedimentary mounds is comparatively limited and/or localized (*4-6*). Quasi-periodic bedding at many locations including the upper portion of Gale's mound implies slow (~30 μm/yr) orbitally-paced accumulation (*10*). These rates are comparable to the modern gross atmospherically-transported sediment deposition rate ($10^{1-2}$ μm/yr ; *20*), suggesting that aeolian processes may be responsible for the layers. Sedimentary strata within Valles Marineris are meters-to-decameters thick, laterally continuous, have horizontal-to-draping layer orientations, and display very few angular unconformities (*11*). These data suggest that sedimentary deposits formed by the accretion of atmospherically-transported sediment (ash, dust, impact ejecta, ice nuclei, or rapidly-saltating sand) were common on both modern and early Mars (*1,9,21*).

Slope-wind erosion of indurated or lithified aeolian deposits cannot explain layer orientations at Gale unless the topographic depression surrounding the mound (i.e., the moat) seen in Figure 1 was present during mound growth. This implies a coupling between mound primary layer orientations, slope winds, and mound relief.

To explore this feedback, we aimed to develop the simplest possible model that can account for the structure and stratigraphy of Mars' equatorial sedimentary rock mounds (see Supplementary Information for details). In one horizontal dimension ($x$), topographic change $dz/dt$ is given by

$$dz/dt = D - E \quad (1)$$

where $D$ is an atmospheric source term and $E(x,t)$ is erosion rate. Initial model topography (Figure 2) is a basalt (nonerodible) crater/canyon with a flat floor of half-width $R$ and long, 20° slopes. To highlight the role of slope winds, we initially assume $D$ is constant and uniform (for example, uniform airfall sediment concentration multiplied by a uniform settling velocity, with little remobilization; *7,12-13*). $E$ typically has a power-law dependence on maximum shear velocity magnitude at the air-sediment interface, $U$:

$$E = k\, U^{\alpha} \quad (2)$$

where $k$ is an erodibility factor that depends on substrate grainsize and induration/cementation (*22,23*), and $\alpha \sim 3\text{-}4$ for sand transport, soil erosion, and rock abrasion (*24*). We assume eroded material does not pile up in the moat but is instead removed from the crater, for example through breakdown to easily-mobilized dust-sized particles (*25*). Shear velocity magnitude is given by

$$U(x) = Uo + \max \left| \int_{x}^{\pm\infty} \frac{\partial z'}{\partial x'} \exp\left( \frac{-|x - x'|}{L} \right) \mathrm{d}x' \right| \quad (3)$$

which is the sum of a background bed shear velocity $U_o$ and the component of shear velocity due to slope winds. The max|±()| operator returns the maximum of downslope (nighttime) or upslope (daytime) winds, $z'$ is local topography, $x$ and $x'$ are distances from the crater center, and $L$ is a slope-wind correlation length scale that represents the effects of inertia. The slope winds are affected by topography throughout the model domain, but are most sensitive to slopes within $L$ of $x$.

Model output characteristically produces Gale-like mound structure and stratigraphy (Figures 2,S3). Katabatic winds flowing down the crater walls inhibit sediment layer accumulation both on the crater walls and for an inertial run-out length on the floor that scales with $L$. Layer accumulation in the quiet crater interior is not inhibited, so layers can be deposited there. Greater wind speeds close to the walls increase sediment erosion and entrainment. The gradient in slope-wind shear velocity causes a corresponding gradient in sediment accumulation, which over time defines a moat and a growing mound. Mound aggradation rate does not change significantly upsection, consistent with observations that show no systematic decrease in layer thickness with height (*21,10*). Growth does not continue indefinitely: when the relief of the mound becomes comparable to that of the crater walls, slope winds induced by the mound itself become strong enough to erode earlier deposits at the toe of the mound. This erosional front steepens the topography and further strengthens winds, so erosion propagates inward from the edge of the mound, leading to a late-stage net erosional state (Supplementary Information).

This evolution does not require any change in external forcing with time; however, simulating discrete, alternating erosional and depositional events with a constant, short characteristic timescale produces the same model output. For the continuous case, the modeled bedding contacts should be interpreted as timelines, whereas for discrete events the bedding contacts represent small-scale unconformities. Exposure of layering at all elevations on the Gale mound show it has entered the late, erosional stage. The mean dip of all sedimentary layers in this radial cut is 4.7°, and erosion progressively destroys the steepest-dipping layers. Exhumed layers are buried to kilometer depths, but relatively briefly, consistent with evidence that clay diagenesis at Gale was minimal (*26*). Alternations in aqueous mineralogy (*3,5,26*) could represent changes in airfall (silicate) input, rather than a change in global environmental chemistry (*27*). During early mound growth, d*z*/dt is not much slower than *D*. If *D* corresponds to vertical dust settling at rates similar to today, then the lower Gale mound accumulated in $10^{7\text{-}8}$ yr, consistent with crater-counts (*28*) and the orbital-forcing interpretation of cyclic bedding (*10*), suggesting that the time represented by the lower Gale mound is a small fraction of Mars' history.

Values of *L* and *D* on Early Mars are not known, but Gale-like shapes and stratigraphy arise for a wide range of reasonable parameters (Figure S3). Consistent with observations across Mars (Figure S4), moats are infilled for small *R/L*, and for the largest *R/L* multiple mounds can develop within a single crater.

*D* could vary on timescales much shorter than the mound growth timescale, for example if Milankovitch cycles pace the availability of liquid water for cementation. To illustrate this, we set $D(t) = D(t{=}0) + D(t{=}0)\cos(nt)$ where $n^{-1} \ll$ mound growth timescale, and show low-angle unconformities can be preserved, with the likelihood of unconformities increasing with elevation and with distance from the mound center (Fig. 2d). In addition, a late-stage drape crosscuts layers within the mound core at a high angle, and is itself broken up by further erosion (Fig. 2d).

Thin mesa units mapped at Gale and more widely on Mars have these characteristics (*2,4*). Deposition at a constant long-term-average rate is unrealistic for the entire mound history because the rate of sedimentary rock formation on Mars is close to zero in the modern epoch (*29*), most likely because atmospheric loss has restricted surface liquid water availability (*27,30*). To explore this, we decreased $D'$ over time; this allows slope winds down the wall to expose layers and form a moat, expanding the portion of parameter space that allows moats and mounds to form.

Our model suggests that in three dimensions, slopes paralleling $U_o$ will erode faster. In contrast to the $U_o$-parallel teardrop shapes observed for unidirectional flow, background-wind/slope-wind interaction should produce elliptical-plan mounds with long axes orthogonal to $U_o$ and to mean yardang orientations. This is consistent with mound orientations at Gale, Nicholson and West Arabia, which are diagnostic for a contribution by slope winds to erosion.

Slope-wind enhanced erosion and transport is incompatible with a deep-groundwater source for early diagenetic cementation of sedimentary rocks at Gale (*3,5,26*), because deep-groundwater-limited evaporite deposition would infill moats and produce near-horizontal strata. A water source at or near the mound surface (ice weathering, snowmelt, or rainfall) is predicted instead (*7,12,27*). Because perennial surface liquid water prevents aeolian erosion, we predict long dry windy periods interspersed by brief wet periods at Gale, similar to observations along the Opportunity rover traverse (*19*). Upon arriving at the mound, MSL can immediately begin to collect observations that will test our model. MSL can confirm aeolian origin using sedimentology measurements, and constrain present-day winds using its meteorology package, past winds by imaging fossilized bedforms, post-depositional tilting by measuring stream-paleoflow directions, and subsurface dissolution using geochemical measurements.

Unconformities, if present, should be oriented away from the center of the present mound and should be more frequent upsection. Gale Crater is geologically diverse (*1-6*), and records many environments including alluvial fans, inverted channels (*4*), and possibly lacustrine sediments at the very bottom of the mound (*6*); however, within the bulk of the mound, slow, orbitally-paced sedimentation and oscillation between reducing and oxidizing conditions could limit preservation of organic carbon.

**Acknowledgements:** We thank Claire Newman, Mark Richardson, Bill Dietrich, Woody Fischer, Mike Mischna, Francis Nimmo, Aymeric Spiga, Dave Stevenson, Oded Aharonson, and especially Katie Stack, for their intellectual contributions. We additionally thank Claire Newman and Mark Richardson for sharing their MarsWRF output for Gale.

**Author contributions:** All authors contributed extensively to the work presented in this paper.

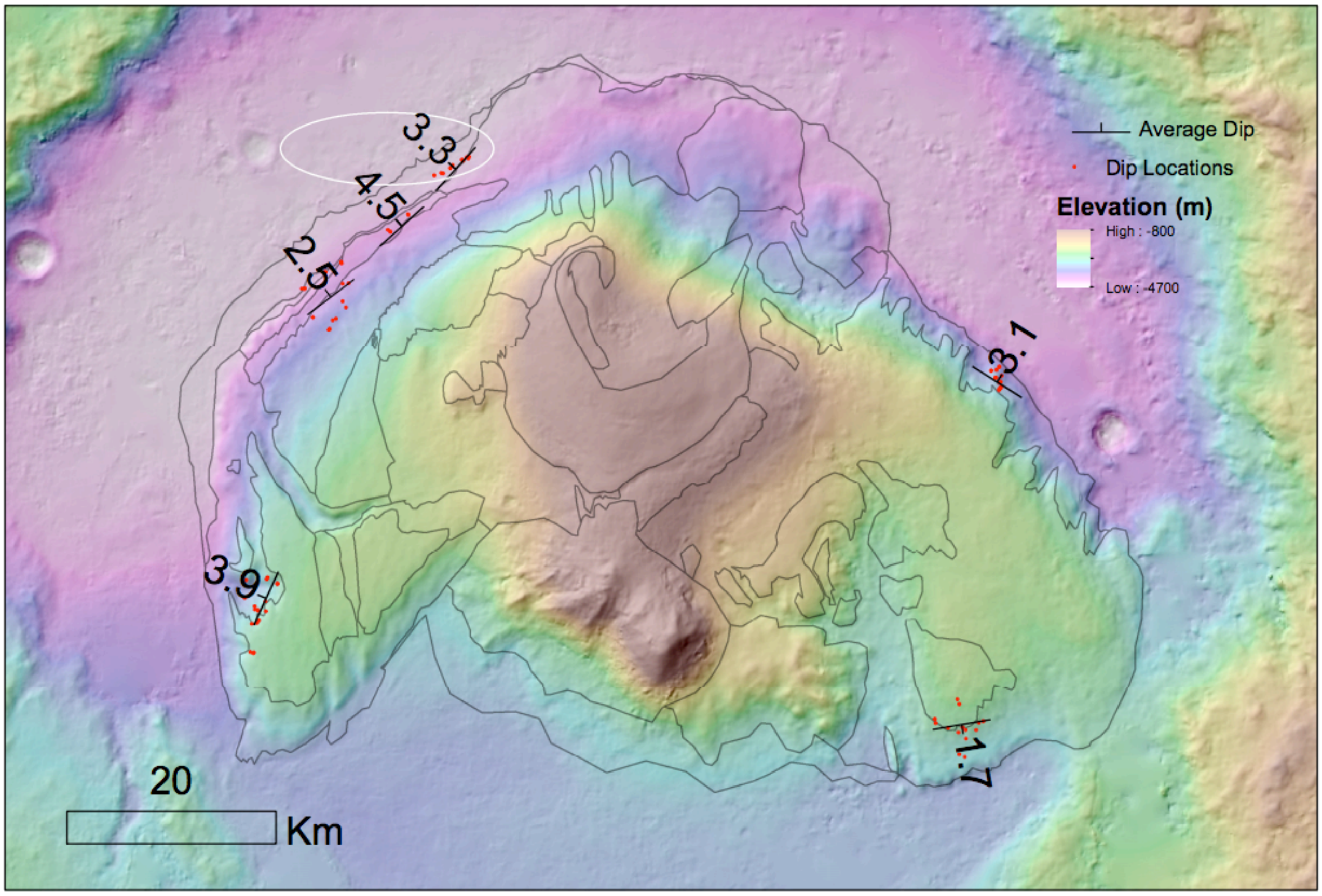

**Fig. 1.** Bedding orientation measurements from six locations around the margin of the Gale crater mound. Individual measurements are marked in red, with the average at each site

indicated by the dip symbol. At each location, beds consistently dip away from the center of the mound, consistent with the proposed model. Background elevation data is from the High-Resolution Stereo Camera (HRSC) (http://europlanet.dlr.de/node/index.php?id=380), with superimposed geologic units from Ref. 5. The MSL landing ellipse is shown in white.

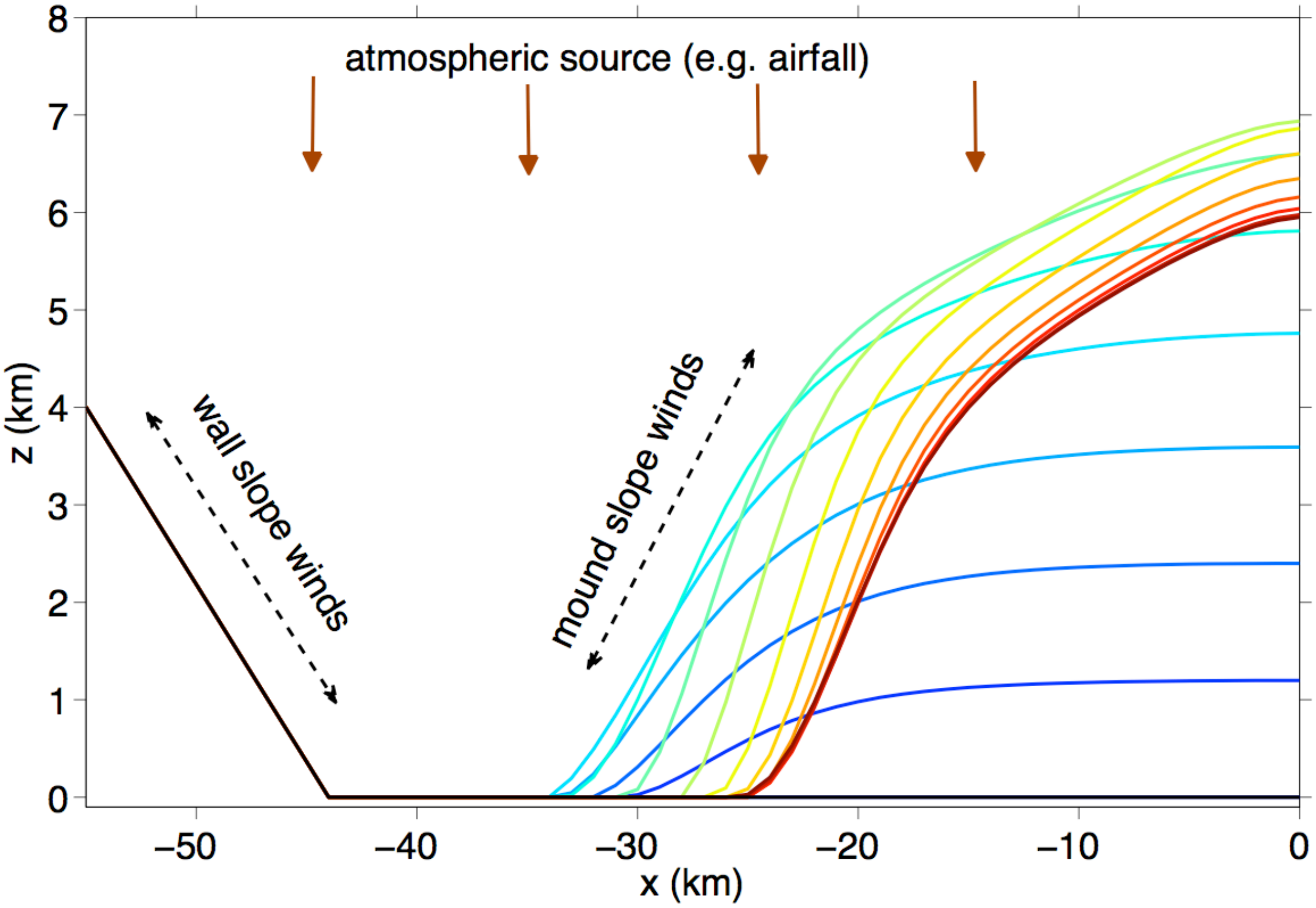

**(a)**

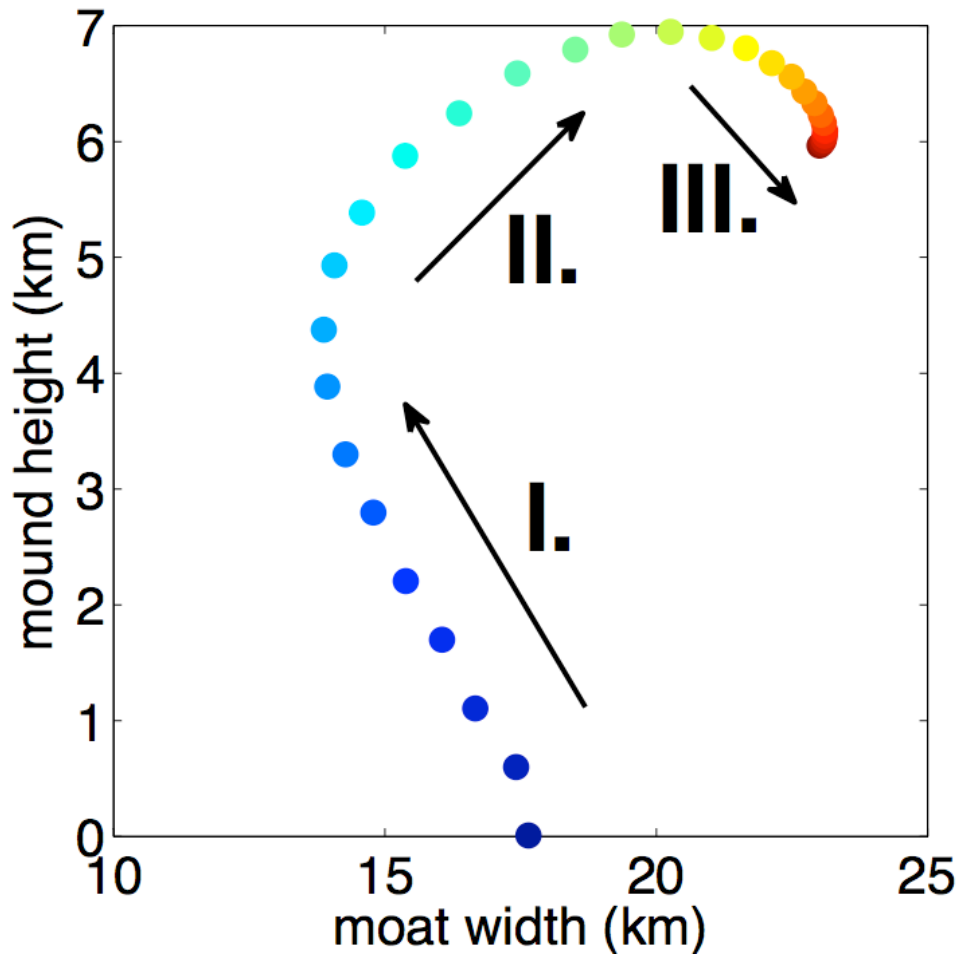

**(b)**

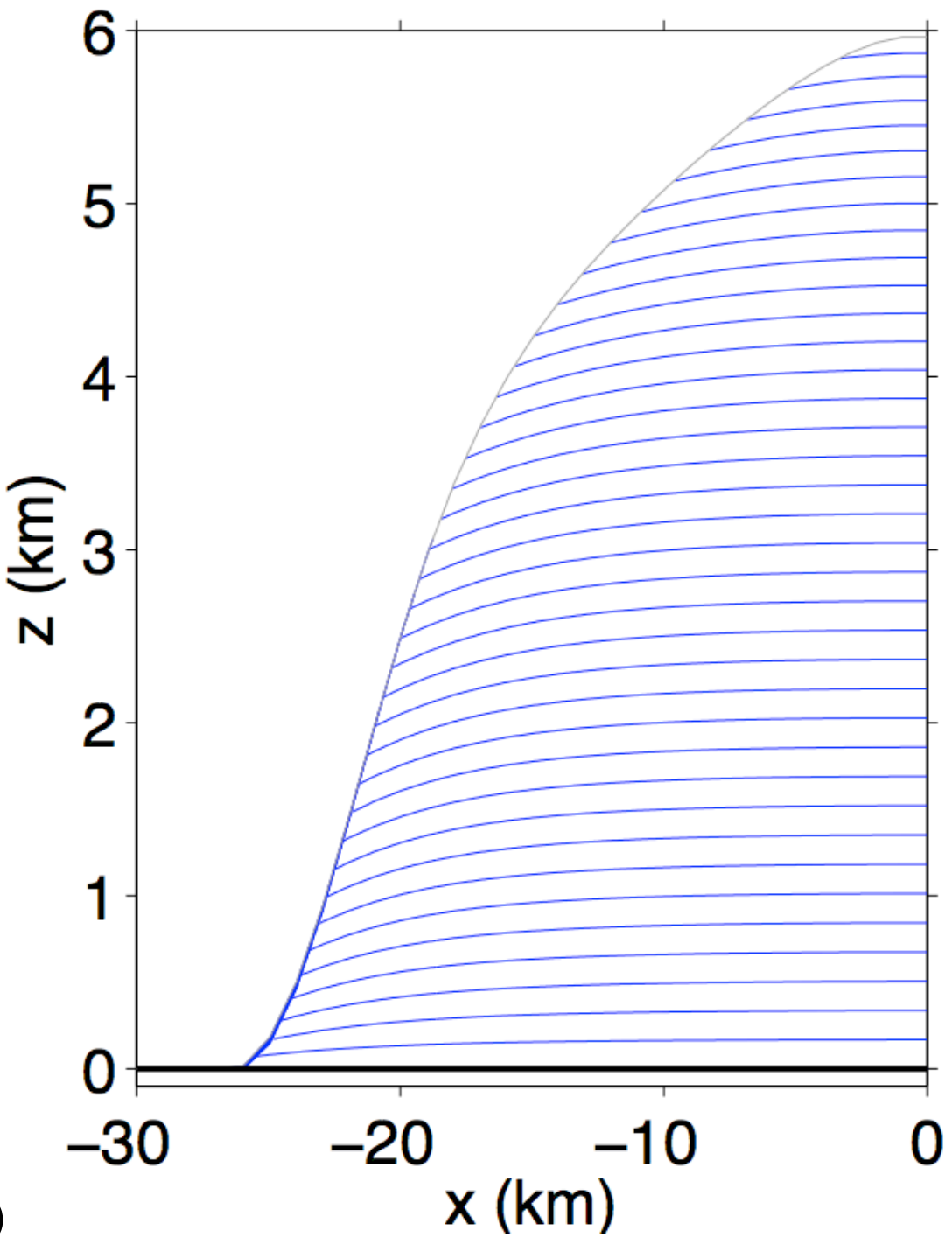

(c)

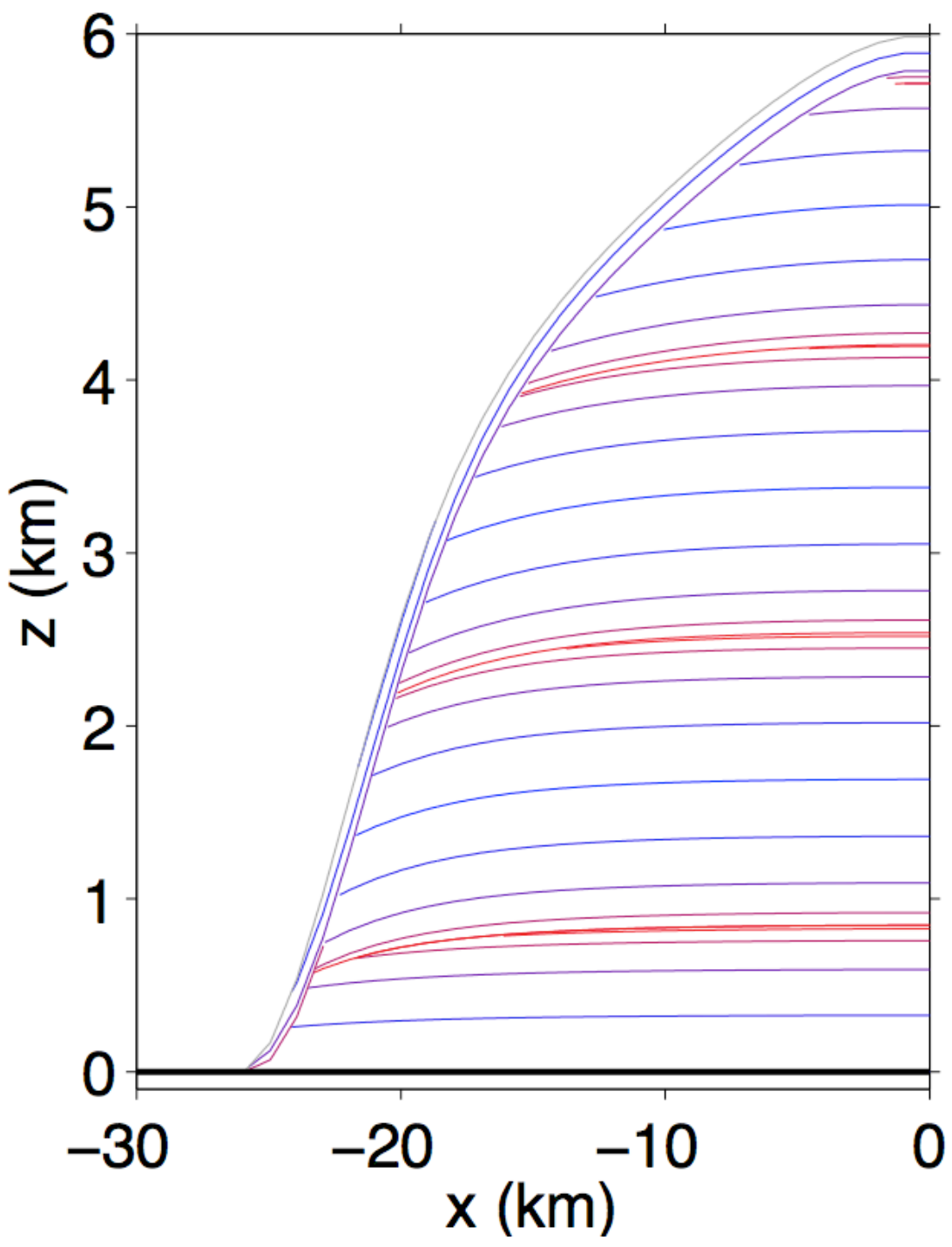

(d)

**Fig. 2.** Simulated sedimentary mound growth and form for $\alpha$=3, *R/L*=2.4, $D'$=0.4. Colored lines in (**a**) correspond to snapshots of the mound surface equally spaced in time (blue being early and red being late), for a radial cut from the crater wall to the crater center. The black line corresponds to the initial topography. *D'* is defined as the deposition rate divided by the mean erosion rate on the crater/canyon floor at simulation start. (**b**) shows mound geometry, with two colored dots for every colored line in (**a**). I, II, and III correspond to phases in the evolution of the mound (Supplementary Text). (**a**) and (**b**) are for steady uniform deposition. Results for time-varying uniform deposition appear very similar at this scale. **(c)** shows stratigraphy formed for steady uniform deposition. Note moatward dips. Flank erosion to form the modern deflation surface (gray) tends to remove any unconformities formed near the edge of the mound, while exposing the stratigraphic record of earlier phases for rover inspection. (**d**) shows stratigraphy resulting from sinusoidally time-varying deposition. For visibility, only a small number of oscillations are shown. Color of strata corresponds to deposition rate: blue is high *D*, which might correspond to wet climates (*27,30*), and red is low *D*, which might correspond to dry climates (*27,30*). Note low-angle unconformities, and late-stage flanking unit intersecting the mound core at a high angle.

## Supplementary information

### 1. Determination of layer orientations

1m-resolution stereo terrain models were produced from High-Resolution Imaging Science Experiment (HiRISE) images, using the method of Ref. 31. To confirm that our procedure is measuring layers within the mound, and is not biased by surficial weathering textures nor by the present-day slope, we made measurements around a small reentrant canyon incised into the SW corner of the Gale mound (Figure S1). Within this canyon, present-day slope dip direction varies through 360°, but as expected the measured layer orientations dip consistently (to the W).

### 2. Assessment of alternative mechanisms for producing outward dips

Few geologic processes can produce primary outward dips of (3±2)° (Figures 1,S2). Spring mounds lack laterally continuous marker beds of the >10 km extent observed (*4*). Differential compaction of porous sediments (*32*), flexural response to the mound load, or flexural response to excavation of material from the moat would tilt layers inwards, contrary to observations. Preferential dissolution, landsliding/halotectonics, post-impact mantle rebound, and lower-crustal flow can produce outward tilting. Preferential dissolution causes overlying rock to fail and leaves karstic depressions (*33*), which are not observed at Gale. Landsliding/halotectonics can produce deformed beds in layered sediments on both Earth and Mars (*34-36*), and a possible late-stage landslide is observed on the Gale mound's north flank (*4*). These sites show order-unity strain and contorted bedding, but the layers near the base of Gale's mound show no evidence for large strains at kilometer scale. On Early Mars, viscoelastic isostatic-compensation timescales are $\ll 10^6$ yr. In order for subsequent mantle rebound to produce outward tilts, the mound must have accumulated at implausibly fast rates. Mars' crust is constrained to be $\lesssim$90 km thick at Gale's location (*37*), so lower-crustal flow beneath 155km-diameter Gale would have a geometry that would relax Gale Crater from the outside in, incompatible with simple outward tilting. Additionally, Gale is incompletely compensated (*38*) and postdates dichotomy-boundary faulting, so Gale postdates the era when Mars' lithosphere was warm and weak enough for limited crustal flow to relax the dichotomy boundary and cause major deformation (*39, 40*). Any tectonic mechanism for the outward dips would correspond to ~3-4 km of floor uplift of originally horizontal layers, comparable to the depth of a fresh crater of this size and inconsistent with the current depth of the S half of the crater if we make the reasonable approximation that wind cannot quickly erode basalt. Tectonic doming would put the mound's upper surface into extension and produce extensional faults (e.g., p.156 in Ref. 41), but these are not observed. In summary, primary dip set by aeolian processes is the simplest explanation for the outward-dipping layers in Gale's mound. MSL can confirm this, for example by comparing stream-deposit paleoflow directions to the modern slope.

**3. Scaling sediment transport**

Conservation of sediment mass (*42*) in an atmospheric boundary-layer column can be written as:

$$\mathrm{d}z/\mathrm{d}t = D - E = CWs - E$$

Here *C* is volumetric sediment concentration, *Ws* is settling velocity, and *E* is the rate of sediment pick-up from the bed. In aeolian transport of dry sand and alluvial-river transport, induration processes are weak or absent and so the bed has negligible intergrain cohesion. *C* tends to *E/Ws* over a saturation length scale that is inversely proportional to *Ws* (for d*z*/d*t* > 0) or *E* (for dz/dt <0). This scale is typically short, e.g. ~1-20m, for the case of a saltating sand on Earth (*24*). Our simplifying assumption that $D \neq f(x)$ and therefore $C \neq f(x)$ implies that this saturation length scale is large compared to the morphodynamic feedback of interest. For the case of net deposition (dz/dt > 0) this could correspond to settling-out of sediment stirred up by dust storms (*45*). These events have characteristic length scales $>10^2$ km, larger than the scale of Gale's mound and justifying the approximation of uniform *D* (*46*). For the case of net erosion (dz/dt < 0), small *E* implies a detachment-limited system where sediments have some cohesive strength (e.g., damp or cemented sediment, bedrock, crust formation). The necessary degree of early induration is not large: for example, 6-10 mg/g chloride salt increases the threshold wind stress for saltation by a factor of *e* (*22*). Shallow diagenetic cementation (*23*) could be driven by snowmelt, rainfall, or fog. Fluid pressure alone cannot abrade the bed, and the gain in entrained-particle mass from particle impact equals the abrasion susceptibility, $\sim 2 \times 10^{-6}$ for basalt under modern Mars conditions (*16*) and generally <<1 for cohesive materials, preventing runaway adjustment of *C* to *E/Ws*. Detachment-limited erosion is clearly appropriate for slope-wind erosion on modern Mars (because sediment mounds form yardangs, shed boulders, and have high thermal inertia), and is probably a better approximation to ancient erosion processes than is transport-limited erosion (given the evidence for ancient near-surface liquid water, shallow diagenesis, and soil crusts) (*23,43,44*).

This makes testable predictions for MSL. For example, the key role attributed to suspended sediment during mound growth predicts that particles too large to be suspended will be uncommon, except as aggregates (*25*). Slope-wind enhanced erosion could, however, contribute to erosion of a mound made of coarse intact grains.

**4. Mound growth dynamics**

Coriolis forces are neglected because almost all sedimentary rock mounds on Mars are equatorial (*27*). Additional numerical diffusivity at the $10^{-3}$ level is used to stabilize the solution. Analytic and experimental results show that in slope-wind dominated landscapes, the strongest winds occur close to the steepest slopes (*47-48*). $L$ will vary across Mars because of changing 3D topography (*49*), and will vary in time because of changing atmospheric density. Ref. 45 shows $L$ ~ 20km for Mars slopes with negligible geostrophic effects. Simulations of gentle Mars slope winds strongly affected by planetary rotation suggest $L$ ~ 50-100 km (*51-52*). Entrainment acts as a drag coefficient with value 0.02-0.05 for Gale-relevant slopes (*47,53-54*), suggesting $L$ = 20-50 km for a 1km-thick cold boundary layer. Dark strips in nighttime thermal infrared mosaics of the horizontal plateaux surrounding the steep-sided Valles Marineris canyons are ~20-50 km wide, which might correspond to the sediment transport correlation length scale for anabatic winds (*18*). Therefore $L \sim 10^{1-2}$ km is reasonable, but with the expectation of significant variability in $L/R$, which we explore in the next section.

Early in mound evolution (Phase I; Figure 2b), mound topography can be a positive feedback on mound growth because the mound's adverse slope decelerates erosive winds flowing down from the canyon walls. The mound toe can either hold position or prograde slightly into the moat, depending on parameter choices. As the mound grows, winds flowing down the mound flanks become progressively more destructive, and a kinematic wave of net erosion propagates inward from the mound toe (Phase II). During the all-erosive Phase III, decreasing the mound height reduces the maximum potential downslope wind. However erosion also decreases mound width, which helps to maintain steep slopes and correspondingly strong winds. Erosion decreases everywhere at late stage, and the model mound can either (i) enter a quasi-steady state where slow continued slope-wind erosion is balanced by diffusive geologic processes such as landsliding, or (ii) reduction in windspeed in the widening moat can lead to cycles of satellite-mound nucleation, autocatalytic growth, inward migration and self-destruction. There is strong evidence for secular climate change on the real Mars, which would break the assumption of constant external forcing (Main Text). $U_o$ is set to zero in Figure 2, and $U_o$ sensitivity tests show that for a given $D'$, varying $U_o$ has little effect on the pattern of erosion because spatial variations are still controlled by slope winds. Equation (3) implies the approximation $E \sim \max(U)^{\alpha} \sim \sum U^{\alpha}$, which is true as $\alpha \rightarrow \infty$. To check that this approximation does not affect conclusions for $\alpha$ = 3-4 (*24*), we ran a parameter sweep with $E \sim (U_+^{\alpha} + U_-^{\alpha})$. For nominal parameters (Figure 2), this leads to only minor changes in mound structure and stratigraphy (e.g., 6% reduction in mound height and <1% in mound width at late time). For the parameter sweep as a whole, the change leads to a slight widening of the regions where the mound does not nucleate or overspills the crater (changing the outcome of 7 out of the 117 cases shown in Figure S3). The approximation would be further supported if (as is likely; *24*) there is a threshold $U$ below which erosion does not occur. If MSL shows that persistent snow or ice is needed as a water source for layer cementation (*7*,*27*), then additional terms will be required to track humidity and the drying effect of föhn winds *(15,55)*.

## 5. Controls on mound growth and form

To determine the effect of parameter choices on sedimentary rock mound size and stratigraphy, we carried out a parameter sweep in $\alpha$, $D'$, and $R/L$ (Figure S3). Weak slope dependence ($\alpha$ = 0.05) is sufficient to produce strata that dip toward the foot of the crater/canyon slope (like a sombrero hat). Similarly weak *negative* slope dependence ($\alpha$ = -0.05) is sufficient to produce concave-up fill.

At low $R/L$ (i.e., small craters) or at low $\alpha$, $D'$ controls overall mound shape and slope winds are unimportant. When $D'$ is high, layers fill the crater; when $D'$ is low, layers do not accumulate. When either $\alpha$ or $R/L$ or both are $\gtrsim 1$, slope-wind enhanced erosion and transport dominates the behavior. Thin layered crater floor deposits form at low $D'$, and large mounds at high $D'$.

If $L$ is approximated as being constant across the planet, then $R/L$ is proportional to crater/canyon size. Moats do not extend to basement for small $R/L$, although there can be a small trench at the break-in slope. For larger $R/L$, moats form, and for the largest craters/canyons, multiple mounds can form at late time because slope winds break up the deposits. This is consistent with data which suggest a maximum length scale for mounds (Figure S4). Small exhumed craters in Meridiani show concentric layering consistent with concave-up dips. Larger craters across Meridiani, together with the north polar ice mounds, show a simple single mound. Gale and Nicholson Craters, together with the smaller Valles Marineris chasmata, show a single mound with an undulating top. The largest canyon system on Mars (Ophir-Candor-Melas) shows multiple mounds per canyon. Steeper crater/canyon walls in the model favor formation of a single mound. Gale-like mounds (with erosion both at the toe and the summit) are most likely for high $R/L$, high $\alpha$, and intermediate $D'$ (high enough for some accumulation, but not so high as to fill the crater) (Figure S3). These sensitivity tests suggest that mounds are a generic outcome of steady uniform deposition modified by slope-wind enhanced erosion and transport for estimated Early Mars parameter values.

**Supplementary References**

**Supplementary Figures**

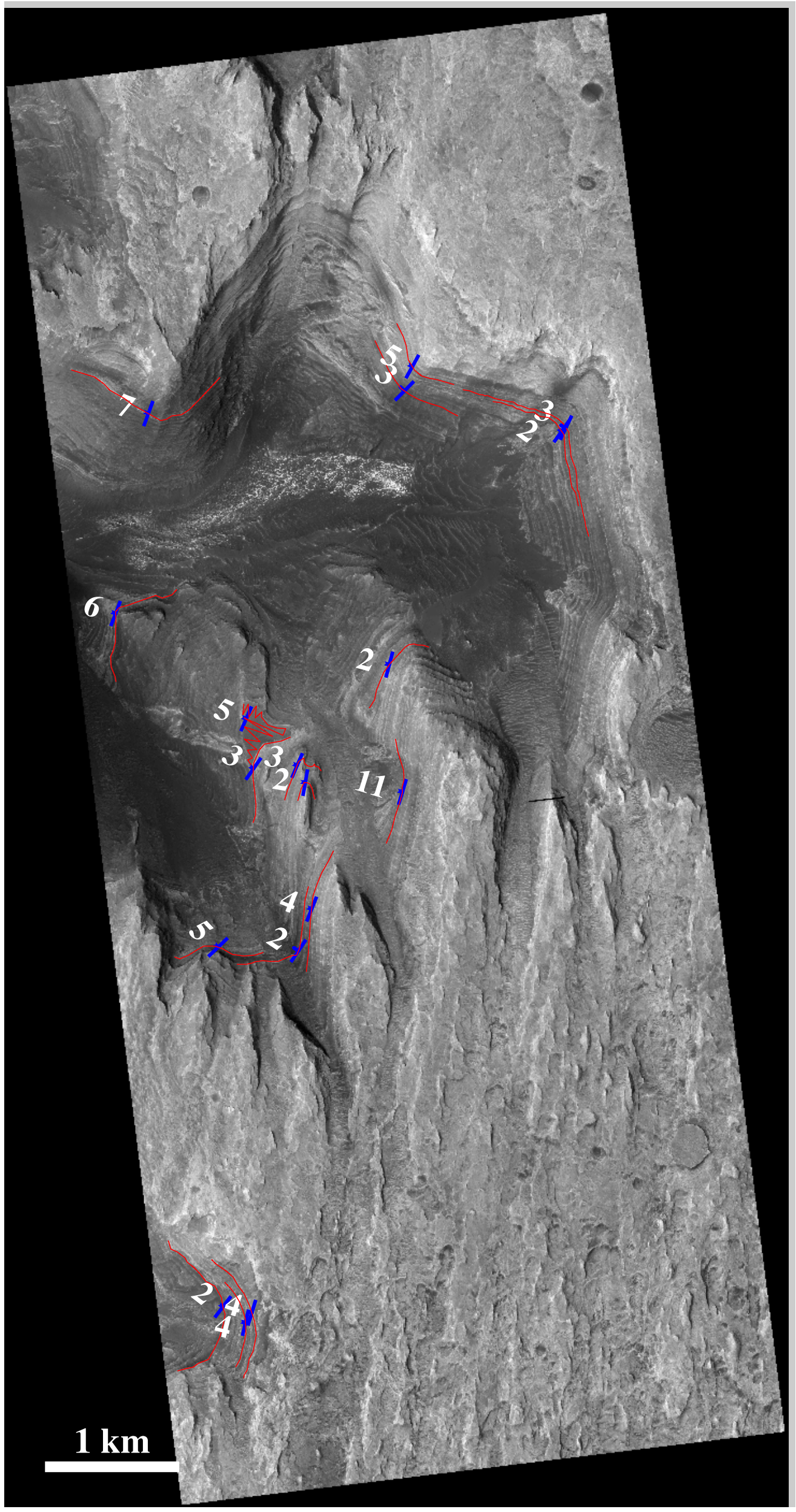

**Figure S1.** Layer orientation measurements from a 1m DTM generated from 25cm/pixel HiRISE stereopair ESP_012907_1745/ESP_013540_1745. This is a small reentrant canyon eroding eastward in the SW part of the mound (the locality in Figure 1 dipping '3.9'). Background is orthoregistered ESP_012907_1745. Red lines are layers traced from images (jagged line corresponds to a planar outcrop). Blue labeled symbols show layer orientations.

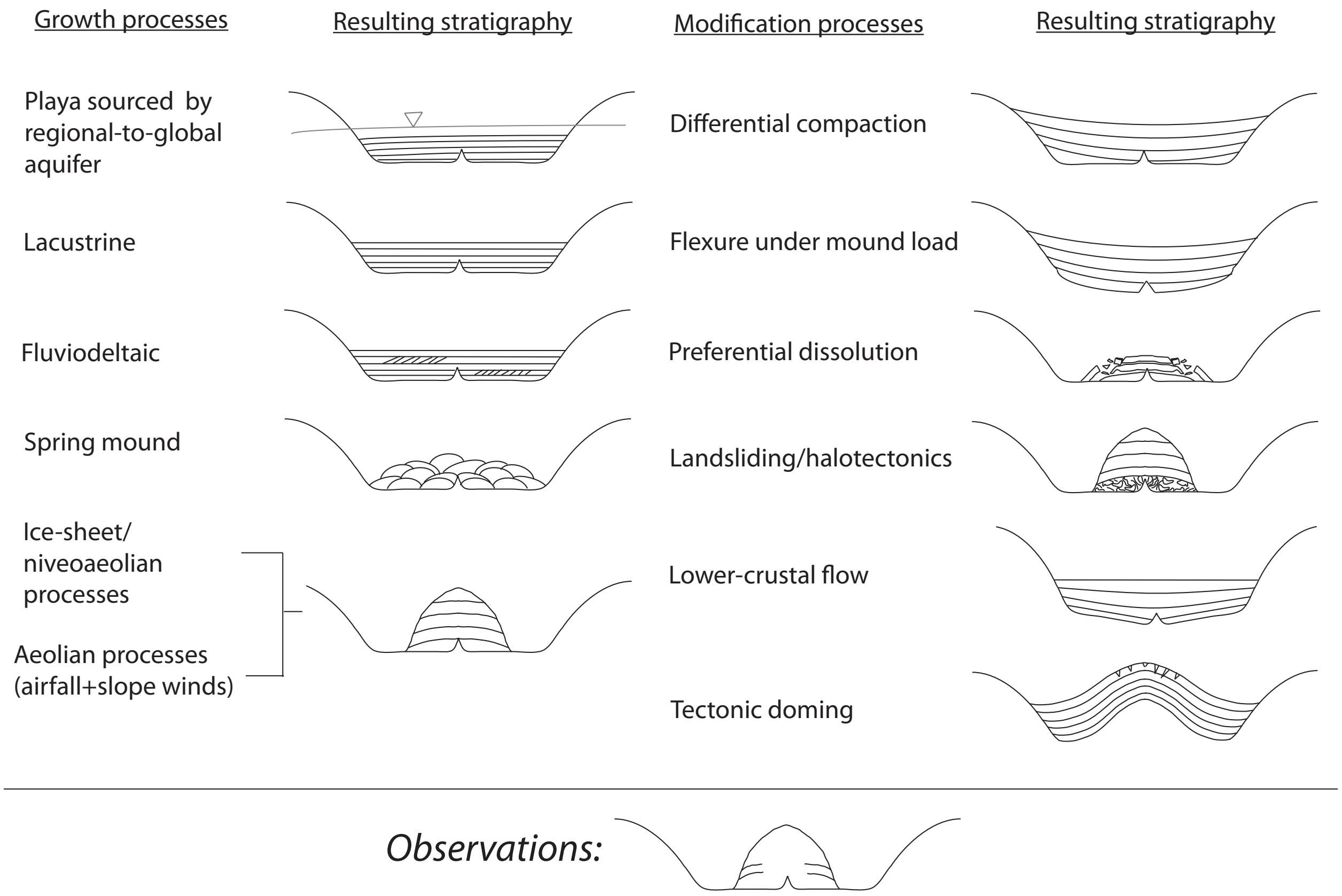

**Figure S2.** Comparison of mound growth hypotheses to measurements, for an idealized cross-section of a mound-bearing crater. Note that groundwater table (gray line highlighted by triangle) does not exactly follow an equipotential (*8*).

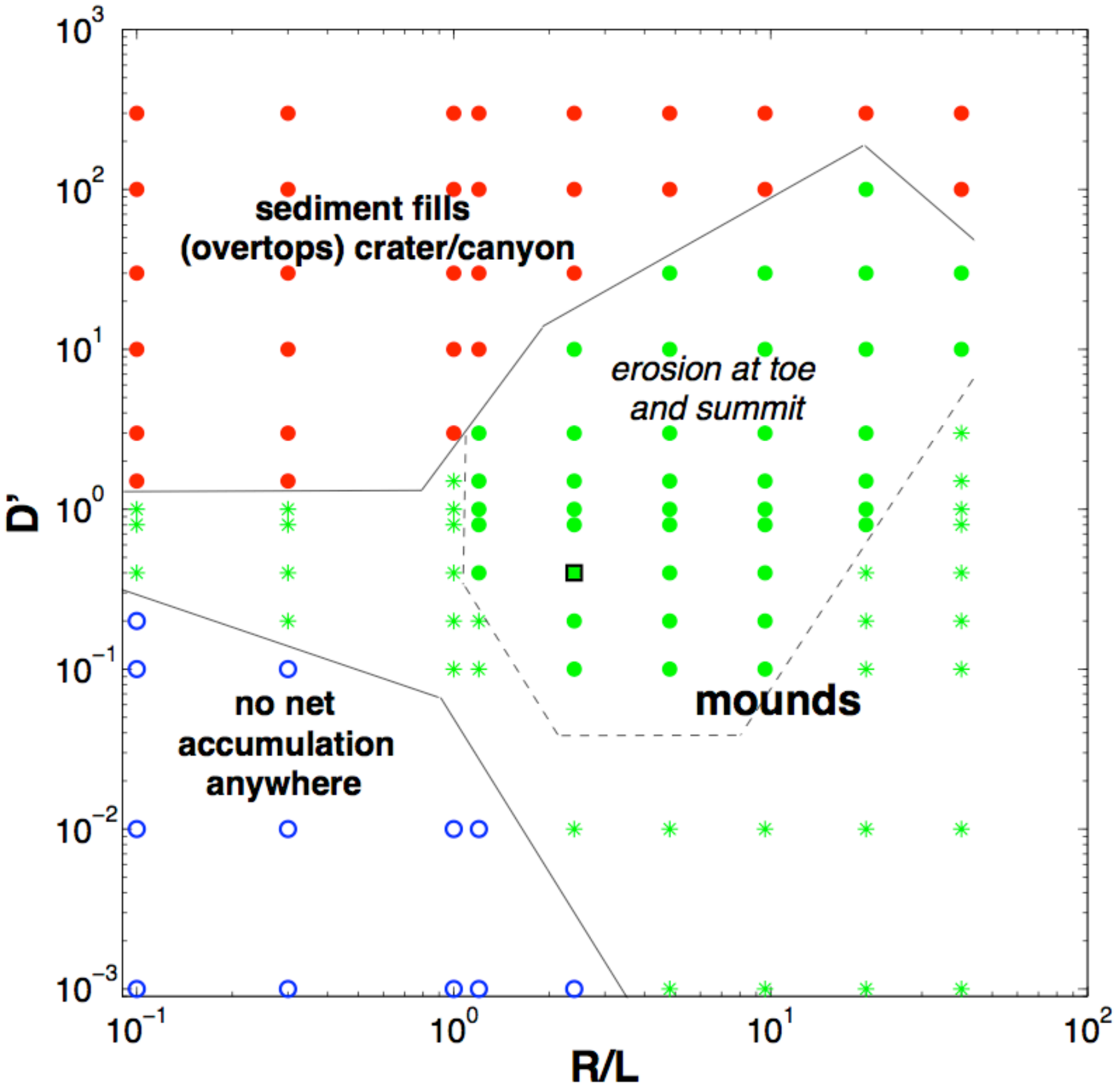

**Figure S3.** Overall growth and form of sedimentary mounds – results from a model parameter sweep varying *R/L* and *D'*, with fixed $\alpha = 3$. Black square corresponds to the results shown in more detail in Figure 2. Symbols correspond to the overall results:– no net accumulation of sediment anywhere (blue open circles); sediment overtops crater/canyon (red filled circles); mound forms and remains within crater (green symbols). Green filled circles correspond to outcomes where layers are exposed at both the toe and the summit of mound, similar to Gale. Multiple mounds form in some of these cases.

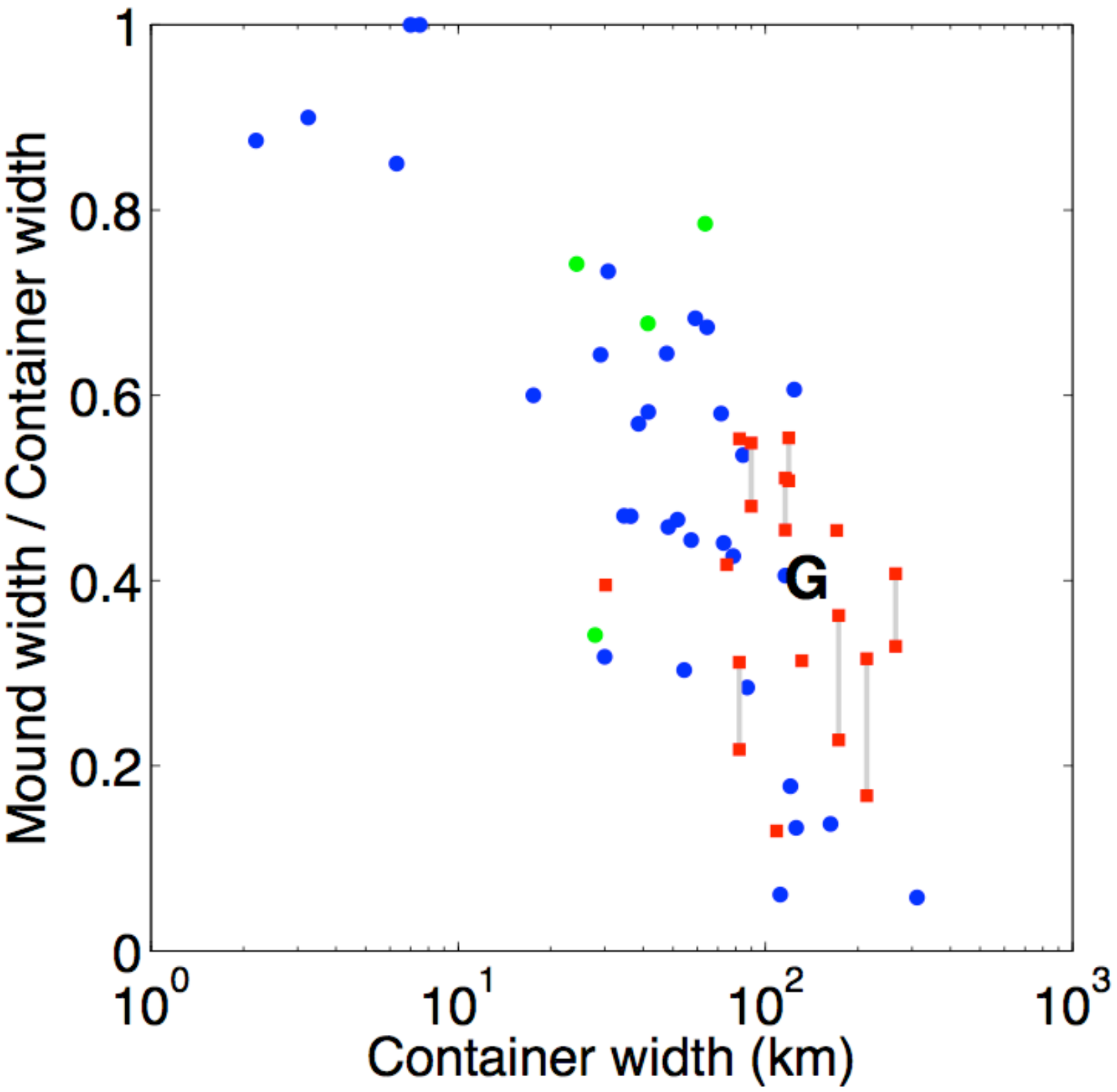

**Figure S4.** Width of largest mound does not keep pace with increasing crater/canyon width, suggesting a length threshold beyond which slope winds break up mounds. Blue dots correspond to nonpolar crater data, red squares correspond to canyon data, and green dots correspond to polar ice mound data. Gray vertical lines show range of uncertainty in largest-mound width for Valles Marineris canyons. Blue dot adjacent to “G” corresponds to Gale Crater. Craters smaller than 10km were measured using Context Camera (CTX) or HiRISE images. All other craters, canyons and mounds were measured using the Thermal Emission Imaging System (THEMIS) global day infrared mosaic on a Mars Orbiter Laser Altimeter (MOLA) base. Width is defined as polygon area divided by the longest straight-line length that can be contained within that polygon.